\documentclass[twocolumn,prl,aps,twocolumn]{revtex4}
\usepackage[latin9]{inputenc}
\usepackage{textcomp}
\usepackage{amsbsy}
\usepackage{amstext}
\usepackage{amssymb}
\usepackage{graphicx}
\usepackage[unicode=true,pdfusetitle,
 bookmarks=false,
 breaklinks=false,pdfborder={0 0 1},backref=false,colorlinks=false]
 {hyperref}
\hypersetup{
 bookmarksnumbered=false,bookmarksopen=false}

\makeatletter

\newcommand{\lyxmathsym}[1]{\ifmmode\begingroup\def\b@ld{bold}
  \text{\ifx\math@version\b@ld\bfseries\fi#1}\endgroup\else#1\fi}
\newcommand{\ket}[1]{|#1\rangle}

\@ifundefined{textcolor}{}
{%
 \definecolor{BLACK}{gray}{0}
 \definecolor{WHITE}{gray}{1}
 \definecolor{RED}{rgb}{1,0,0}
 \definecolor{GREEN}{rgb}{0,1,0}
 \definecolor{BLUE}{rgb}{0,0,1}
 \definecolor{CYAN}{cmyk}{1,0,0,0}
 \definecolor{MAGENTA}{cmyk}{0,1,0,0}
 \definecolor{YELLOW}{cmyk}{0,0,1,0}
}

\usepackage{amsmath}
\usepackage{graphicx}
\usepackage{amssymb}
\usepackage{txfonts,color}

\makeatother

\begin{document}

\title{Sensing in the Presence of Observed Environments}

\author{Martin B. Plenio}
\email{martin.plenio@uni-ulm.de}
\affiliation{Institut f\"ur Theoretische Physik and IQST, Albert-Einstein-Allee 11, Universit\"at
Ulm, D-89069 Ulm, Germany}

\author{Susana F. Huelga}
\affiliation{Institut f\"ur Theoretische Physik and IQST, Albert-Einstein-Allee 11, Universit\"at
Ulm, D-89069 Ulm, Germany}

\begin{abstract}
    Sensing in the presence of environmental noise is a problem of increasing practical interest.
    In a master equation description, where the state of the environment is unobserved, the
    effect of signal and noise is described by system operators only. In this context it is well-known
    that noise that is orthogonal on an external signal can be corrected for without perturbing the signal,
    while similarly efficient strategies for non-orthogonal signal and noise operators are not known.
    Here we make use of the fact that system-environment interaction typically arises via local two-body
    interactions describing the exchange of quanta between system and environment, which are observable
    in principle. That two-body-interactions are usually orthogonal on system operators,
    allows us to develop error corrected sensing supported by the observation of the quanta that are
    emitted into the environment. We describe such schemes and outline a realistic proof-of-principle
    experiment in an ion trap set-up.
\end{abstract}
\maketitle

{\em Introduction --} The use of quantum resources to improve sensing and metrology
has a longstanding history that originated in proposals for using single mode squeezed
states \cite{Caves1981} and multi-particle spin-squeezing, i.e. entanglement, in
precision measurements \cite{WinelandBI+1992,WinelandBI+1994,GiovannettiLM2011}.
Later it was recognised that environmental noise has a non-trivial impact on the
performance of entanglement-based schemes \cite{HuelgaMP+97}. Given a fixed
number of particles $n$ and a total available time $T$ for the measurement to be
completed, uncorrelated and maximally entangled preparations of $n$ particles were
shown to achieve exactly the same precision for phase measurements when subject to
Markovian dephasing. That is, noise parallel to the signal and described by a time
homogeneous (Lindblad) master equation. The precision of phase estimation subject
to this type of noise becomes (asymptotically) standard quantum limited ($\sim 1/\sqrt{T}$)
and only a constant ($n$-independent) improvement over the shot noise limit is
achievable by means of certain partially entangled initial state preparations
\cite{HuelgaMP+97}. The validity of this result for arbitrary measurements \cite{escher}
and local quantum operations beyond dephasing \cite{rafal} has now been rigourously proven.
This metrological equivalence of product and entangled states fails though in the presence
of non-Markovian (non Lindbland) noise \cite{ChinHP2012,simon}, spatially correlated
noise \cite{JeskeCH2014,dorner} and noise acting on a preferred direction \cite{ChavesBM+2013}.

Recognising the importance of Markovian noise in entanglement based sensing and
precision spectroscopy, the use of methods from quantum error correction was also
explored. Symmetrization procedures lead to some improvements for phase estimation
in the presence of local dephasing noise \cite{HuelgaMC+2000}. Furthermore, it was
also pointed out that quantum error correction can be used to correct the effects
of noise without affecting the signal in cases where the Lindblad operators $L$
describing the environmental noise are orthogonal on the signal operators $S$, in
the sense that conditions of the form $\mbox{tr}[LS^{\dagger}]=0$ are satisfied
\cite{Preskill2000}. This result utilize the principles
underlying the application of quantum gates on quantum error correction codes in fault
tolerant quantum computation (see \cite{DevittMN2013} for an introduction). The use of error correction
has received renewed attention recently \cite{ArradVA+14,KesslerLS+2014,DurSF+14,Ozeri2013,HerreraAR2014}
in particular highlighting the fact that in the presence of local noise, interaction parameters
corresponding to many-body coupling terms can, in principle, be determined up to the
Heisenberg limit ($\sim 1/T$).

The question remained open however as to whether quantum error correction might be able
to achieve similar improvements, allowing to reach the Heisenberg limit, in the presence
of non-orthogonal noise, that is, when the Lindblad operator $L$ that describes the environmental noise and the
signal operator $S$ satisfy $\mbox{tr}[LS^{\dagger}]\neq 0$. While \cite{HuelgaMC+2000} was able
to find some sensitivity improvements for parallel noise, where noise and signal operators
are actually identical, the goal of reaching Heisenberg scaling remained elusive.

\begin{figure}[!h]
\includegraphics[width=8.5cm]{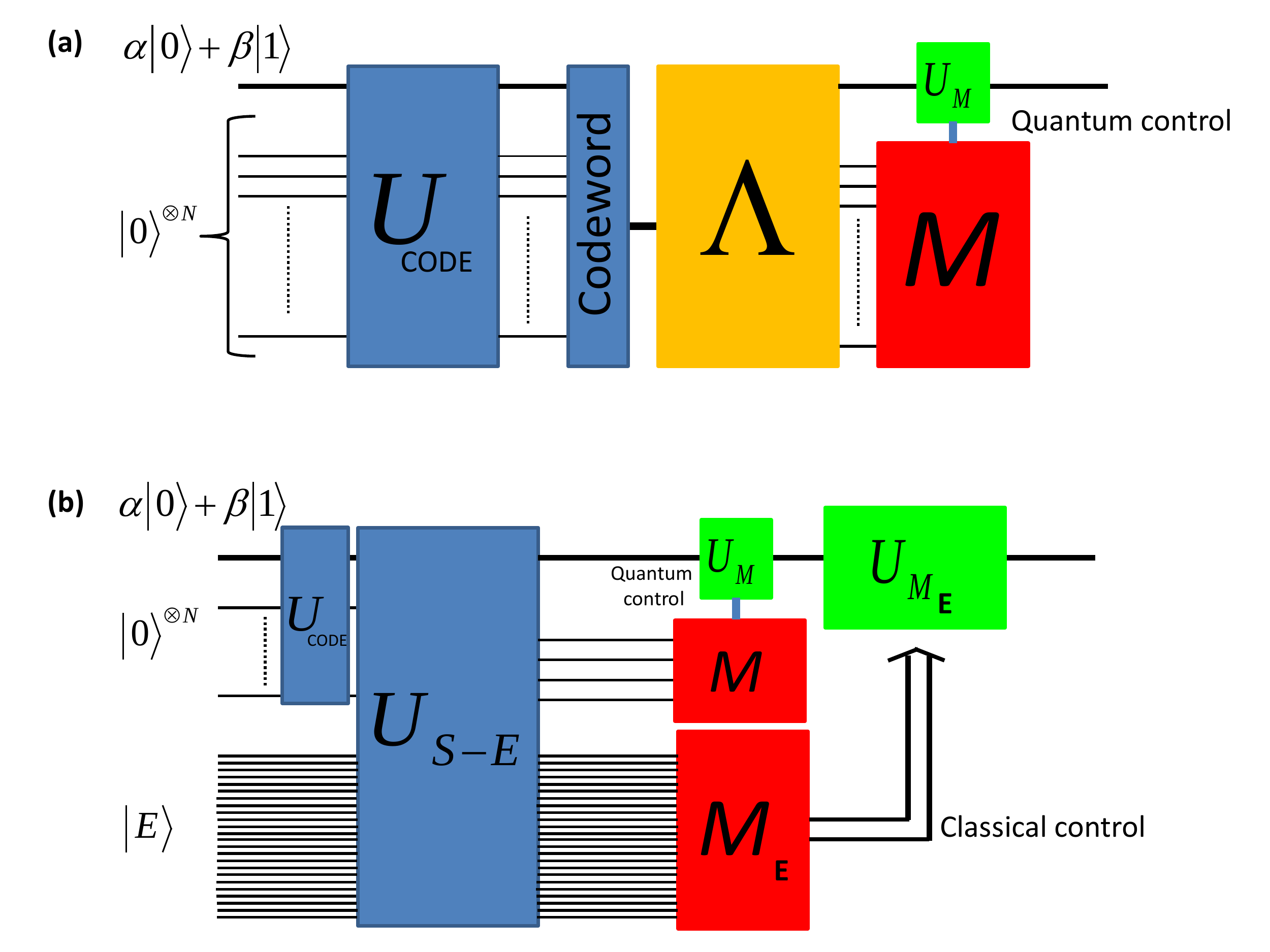}
\caption{Observation of some environmental degrees of freedom can facilitate the persistence of
quantum advantage for sensing and metrology. (a) Idealized quantum error correction. The qubit
sensor, initially in an arbitrary state $\ket{\psi} \alpha \ket{0}+\beta \ket{1}$, is unitarily
manipulated together with $N$ ancillary systems in some reference state $\ket{0}^{\otimes N}$ so
that to generate a higher dimensional system (codeword) which is exposed to the action of a quantum
channel $\Lambda$ modelling the noisy evolution of a desired qubit unitary $U$. Subsequently
a quantum operation coherently controlled by the state of the error syndrome then performs a
suitable correction step $U_M$ on the probe. (b) Environmental monitoring enlarges the ancillary
system and permits the retrieval of some information leaked out of the system probe and into the
environment in order to complement the coherent error on the basis of the syndrome state by
means of a unitary operation conditional on the classical information obtained from the environment.}
\label{Figure1}
\end{figure}

It is important to notice that the considered approaches all make use of the information
that is effectively contained in the system
alone and ignore the information that has leaked out into the environment. In some cases however,
it can be possible to obtain partial information about the state of the environment with
which the system is interacting. This additional information may supplement that obtained
from the measurements performed on the error correction code and may therefore allow for
schemes that are not possible within the standard formulation of quantum error correction.

Such an approach can be interpreted as an extended error correction code in which, due to
the lack of control of the environment, we do not have access to the full information about
the error syndrome \cite{DevittMN2013} and may not perform quantum operations on the system
qubit that are coherently controlled by the syndrome qubits. The best that one can hope for
here is to perform measurements that generate partial information about the environment state
followed by some classically controlled unitary operation on the sensing qubit, as schematically illustrated in figure 1.

In this work we show that this formalism allows for the correction of spontaneous emission,
and more generally for system-environment interactions in which the environment operators are
traceless, while acquiring signals accumulated by certain types of single qubit operators. We
also discuss a possible realisation which overcomes the challenges imposed
by the requirement of the high efficiency detection of all the quanta that are generated in
the environment.

{\em Error corrected sensing of amplitude signals for observed environments --}
In order to bring out the basic principles of the proposed approach we consider the
paradigmatic case of spontaneous decay. The dynamics of spontaneous decay of a qubit (basis states
$|0\rangle,|1\rangle$) into a broad band environment of bosonic degrees of freedom with
bosonic operators $a_{k}$ is described, within rotating wave approximation, by the Hamiltonian
\begin{equation}
    H_{S-E} = \sum_{k} g_k (\sigma_+ a_k +  \sigma_- a_k^{\dagger}).
\end{equation}
As this Hamiltonian is orthogonal on single body operators in the system alone it should
be possible to correct for the effect of its action if full control over the environment
was available. This is due to the fact that in this case the problem reduces to that of
ordinary quantum error correction where the signal is a single body operator which is
orthogonal in the noise which is a two-body operator for which error correction strategies
are known \cite{Preskill2000,ArradVA+14,KesslerLS+2014,DurSF+14,Ozeri2013,HerreraAR2014}.
Such control would allow us to apply recovery operators conditional on the state of the
environment. This is extraordinarily challenging and usually impossible however. It is
this lack of full control over the environmental degrees of freedom which prevents us from taking this direct
route. Instead we will have to devise a scheme which employs classical information obtained
in measurements on the environment and uses this information to apply corrective unitary
operations on the system.

Let us first consider the state of system and environment without the signal Hamiltonian.
Together with the environment Hamiltonian $H_E = \sum_k \hbar\omega_k a_k^{\dagger}a_k$ and
after an evolution time $\Delta t$ that well exceeds the correlation time of the environment,
we find that an initial S-E product state, where the qubit system is prepared in an arbitrary 
superposition state and the environment is initially in the vacuum state, will evolve into an entangled state
of the form \cite{PlenioK1998}
\begin{equation}
    (\alpha|0\rangle + \beta|1\rangle)|0_{\gamma}\rangle \rightarrow (\alpha|0\rangle + \beta e^{-\gamma \Delta t}|1\rangle|0_{\gamma}\rangle + \beta\sqrt{1-e^{-2\gamma \Delta t}}|0\rangle|\phi_{\gamma}\rangle
\end{equation}
where $|\phi_{\gamma}\rangle$ is the state of the environment after the time $\Delta t$ and
$\gamma$ is the effective excitation decay rate of the system. If the aim were to preserve the initial qubit superposition intact, we see that two errors occur: (i) If a quantum has been generated
in the environment then the system moves from the excited state to the ground state and (ii) if
no quantum has been generated the weight of the excited state is reduced.

We can compensate the error originating from (ii) by applying in regular time intervals $\Delta/2$
a $\pi$-pulse on the system qubit which for a time interval $[0,\Delta t]$ would result in the
mapping
\begin{eqnarray}
    (\alpha|0\rangle + \beta|1\rangle)|0_{\gamma}\rangle &\rightarrow& \\
    && \hspace*{-1.5cm} e^{-\gamma\Delta t/2}
    (\alpha|0\rangle + \beta |1\rangle|0_{\gamma}\rangle + \sqrt{1-e^{-\gamma\Delta t}}
    |\psi\rangle|\phi_{\gamma}\rangle \nonumber
\end{eqnarray}
where $|\psi\rangle = |0\rangle + O(\Delta t)$ due to the application of the pulses on the
state that has already suffered a spontaneous decay. As we will observe later the $O(\Delta t)$
correction is negligible in the overall error budget and the limit $\Delta t\rightarrow 0$.
As the time of the spontaneous decay is not known we must chose $\Delta t \ll \gamma^{-1}$
which requires the rapid application of the $\pi$-pulses and therefore a considerable amount
of power and total energy, the latter scaling at $1/\Delta t$.

Alternatively, using knowledge of the decay rate $\gamma$ which can be obtained from a careful
characterisation of the system as well as knowledge of the choice of $\alpha$ and $\beta$,
we can compensate for the error (ii) either by a short laser pulse at time $\Delta t$ that
achieves $(\alpha|0\rangle + \beta e^{-\gamma \Delta t}|1\rangle \rightarrow \alpha|0\rangle
+ \beta|1\rangle$ or via the continuous application of a compensating driving field using the
Hamiltonian
\begin{equation} \label{compensate}
    H_{com} = -\gamma\alpha\beta \sigma_y
\end{equation}
such that we find
\begin{eqnarray}
    (\alpha|0\rangle + \beta|1\rangle)|0_{\gamma}\rangle &\rightarrow& \\
    && \hspace*{-1.5cm} e^{-\gamma\beta^2 \Delta t}
    (\alpha|0\rangle + \beta |1\rangle|0_{\gamma}\rangle + \sqrt{1-e^{-2\gamma\beta^2 \Delta t}}
    |\psi'\rangle|\phi_{\gamma}'\rangle \nonumber
\end{eqnarray}
where $|\psi'\rangle = |0\rangle + O(\Delta t)$ differs from the ground state due to the
dynamics of the compensating Hamiltonian $H_{com}$. Here the driving field intensity is
independent of $\Delta t$.

The first approach requires accurate pulse timing and is state independent, while the second
approach requires accurate knowledge of the decay rate and stability of the driving field.
Which approach is preferable depends on the experimental setting.

Now we note that the two situations (i) and (ii) correspond to states of the environment that
are orthogonal and easily distinguished without the necessity of obtaining the full information
about $|\phi_{\gamma}\rangle$. In case (i) the environment is in a state that is a superposition
of single- or multiquantum states in (ii) the environment is strictly in the ground state. Hence, instead
of applying a recovery operation conditional on the full state of the environment it will be sufficient
to act conditional on the number of quanta in the environment. This does not require the coherent
control of the action on the system by the environment but can be achieved by first measuring the
excitation number in the environment and the subsequent application of a quantum gate on the system
degrees of freedom. Naturally, the recovery cannot be achieved with just a single system qubit and
additional code qubits are required as we will discuss now.

We have not been able to identify a strategy that is applicable for a general signal operator. Instead,
here we present a procedure that is capable of error corrected sensing for specific signal operators
representing rotations about an axis in the $x-y$ plane. For simplicity we begin by considering
\begin{equation}
    H_{sig} = g\sigma_x,
\end{equation}
a case that is easily generalised to the more general $H_{sig} = g(\cos\phi \sigma_x + sin\phi\sigma_y)$.
Following \cite{ArradVA+14} we chose a code that is composed of a sensing qubit subject to error (e.g.
the electron spin of an NV center in diamond) and a second robust qubit (e.g. a nearby carbon nuclear spin)
that is assumed to be isolated from the environment. This assumption is not limiting generality as
there are error correcting codes for spontaneous emission \cite{PlenioVK1997,LeungNC+1997} and general
errors \cite{DevittMN2013} which can emulate the robust qubit by means of a larger number of fragile
qubits as long as these qubits are not themselves subjected to the signal Hamiltonian.

Now we chose the code words
\begin{eqnarray*}
    |0_L\rangle &=& \frac{1}{\sqrt{2}}(|0\rangle + |1\rangle) |0\rangle \\
    |1_L\rangle &=& \frac{1}{\sqrt{2}}(|0\rangle - |1\rangle) |1\rangle
\end{eqnarray*}
which are eigenstates to the signal Hamiltonian. We can again correct for the error of type (i)
by the application of $\pi$-pulses after time intervals $\Delta t/2$ or as the two code words differ
merely in the relative phase via the application of a state dependent compensating Hamiltonian $H_{com}
= -\frac{\gamma}{2}\sigma_y\otimes |0\rangle\langle 0| + \frac{\gamma}{2}\sigma_y\otimes |1\rangle\langle 1|$.
In both cases we obtain after a short time interval $[t,t+\Delta t]$ the mapping
\begin{eqnarray*}
    (e^{igt}|0_L\rangle + e^{-igt}|1_L\rangle)|0_{\gamma}\rangle &\rightarrow& \\
    && \hspace{-1.5cm}
    e^{-\gamma\Delta t/2}(e^{ig(t+\Delta t)}|0_L\rangle + e^{-ig(t+\Delta t)}|1_L\rangle)|0_{\gamma}\rangle\\
    && \hspace{-1.5cm} + U_{res}(\Delta t)|0\rangle(e^{ig(t+\Delta t)}|0\rangle - e^{-ig(t+\Delta t)}|1\rangle)|\phi_{\gamma}\rangle
\end{eqnarray*}
where $U_{res}(\Delta t)$, acting on both qubits and thus corrupting the signal, accounts for
corrections due to the dynamics induced by $H_{sig} + H_{com} + H_{S-E}$. We will demonstrate
later that this effect becomes negligible for $\Delta t \rightarrow 0$. Now observation of
the environment allows us to implement a correcting operation on the encoded qubits alone.
Detecting the absence a photon, no action is required while in the case of the detection of
a photon we re-prepare the first sensing qubit in the standard state $|0\rangle$ e.g. via
optical pumping, followed by the application of the two-body unitary operator
\begin{equation}
    V = |0_L\rangle\langle 00| - |1_L\rangle\langle 01|.
\end{equation}
This results in the state after the error correction step
\begin{eqnarray}
    (e^{ig(t+\Delta t)}|0_L\rangle + e^{-ig(t+\Delta t)}|1_L\rangle)|0_{\gamma}\rangle + O(\Delta t)
\end{eqnarray}
where we have assumed that the environment has been returned to the ground state
and account for residual small errors in the state which arise {\em only} in the
case that an excitation has been detected in the environment. A brief estimate
confirms that the corrections can be made negligibly small in the limit $\Delta t\rightarrow 0$.
We perform an error correction step every $\Delta t$ and therefore carry out $N=T/\Delta t$
repetitions in a time interval $[0,T]$. In each time interval of length $\Delta t$ the probability
for the detection of an excitation in the environment is proportional to $\gamma\Delta t$.
If such an excitation has occurred, then a single error correction step will lead to a residual
error in the probability amplitude of the state due to $U_{res}(\Delta t)$ which scales
as $\gamma\Delta t$ so that the overall error in one error correction step scales
as $\gamma^2\Delta t^2$. For a total $N=T/\Delta t$ of repetitions and under
the pessimistic assumption that the errors all add up in phase so that the total
amplitude error scales as $T\gamma^2\Delta t$ and thus vanishes in the limit
$\Delta t\rightarrow 0$.

It is straightforward to extend this approach to any signal Hamiltonian of the form
$H = g (\cos\phi \sigma_x + \sin\phi \sigma_y)$ as this merely requires a rotation in
the x-y plane of the code to
\begin{eqnarray*}
    |0_L\rangle &=& \frac{1}{\sqrt{2}}(|0\rangle + e^{i\phi}|1\rangle) |0\rangle \\
    |1_L\rangle &=& \frac{1}{\sqrt{2}}(|0\rangle - e^{i\phi}|1\rangle) |1\rangle
\end{eqnarray*}
and an adjustment of the compensating Hamiltonian to $H_{com} = -\frac{\gamma}{2}(-\sin\phi \sigma_x
+ \cos\phi \sigma_y)\otimes |0\rangle\langle 0| + \frac{\gamma}{2}(-\sin\phi \sigma_x
+ \cos\phi \sigma_y)\otimes |1\rangle\langle 1|$. Remarkably, for a signal Hamiltonian
of the form $H_{sig} = \omega\sigma_z$ we encounter the problem, that it would require
code state $|0_L\rangle = |0\rangle|0\rangle$ and $|1_L\rangle = |1\rangle |1\rangle$.
In this case, the application of $\pi$-pulses would however lead to the averaging of
the signal and therefore the failure of this approach. Whether our inability to find
a scheme for $\sigma_z$ signals represents a lack of imagination or is due to a fundamental
impossibility represents an open question. It should be noted however, that results
in \cite{DemkoviczM2014} point towards the latter.

{\em Implementations --} The assumption that the environment can be observed continuously
and with near perfect efficiency is highly challenging in general. This is a central problem
in any possible application as the proposed scheme suffers a large error when even a single
excitation in the environment is missed. Hence, the detection of around $99\%$ of all emitted
excitations would be required ensure that merely a $1\%$ residual error is incurred in each
error correction step and therefore extend the coherence by around two order of magnitude.
Typically, photodetectors do not have perfect quantum efficiency and furthermore, it is not
straightforward to observe the full solid angle of the environment. This makes the practical
realisation of the scheme challenging. It should be noted however that in an important setting
for precision sensing, namely trapped ions, methods for addressing this problem have already
been developed. Indeed, \cite{HempelLJ+2013,WanGW+2014} have proposed methods for the detection
of individual photon scattering events by their recoil on the emitting ion. As suggested in
\cite{HempelLJ+2013} near unit efficiency can be achievable by making use of several vibrational
modes. Further improvements can be achieved by combining this scheme with standard photodetectors
that observe the emitted light directly and cavities to favour specific emission directions.

Furthermore, for trapped atoms or defect centers in diamond high detection efficiencies may
be obtained by placing the emitters inside an optical fibre to ensure directed emission into
a high efficiency detector.

{\em Conclusions --} In this note we have demonstrated that the information obtained
by observing an environment that leads to spontaneous decay of a sensing qubit can
be used to allow for error correction. Importantly, correction is possible while at the same time permitting the accumulation of a
signal emerging from a Hamiltonian that is not orthogonal on the Lindblad operators that
describe the spontaneous emission noise. The present scheme makes use of the fact that
quanta in the environment can be detected and in this manner errors that are non-orthogonal, and
hence not distinguishable by standard error correction approaches, can still be treated. This feature
is not specific to spontaneous emission and the generalisation to other system-environment interactions
follows directly from the principle shown here. A possible experimental realisation of high
efficiency detection of scattered photons in an ion trap has been alluded to.

{\em Acknowledgements --} This work was supported by the Alexander von Humboldt Foundation,
the ERC Synergy grant BioQ, the EU projects DIADEMS, QUCHIP and SIQS. We acknowledge discussions
with Alex Retzker.

\end{document}